# On-demand generation of shallow silicon vacancy in silicon carbide

Jun-Feng Wang[1, 2], Qiang Li[1, 2], Fei-Fei Yan[1, 2], He Liu[1, 2], Guo-Ping Guo[1, 2], Wei-Ping Zhang[3], Xiong Zhou[3], Zhi-Hai Lin[1], Jin-Ming Cui[1,2], Xiao-Ye Xu[1,2], Jin-Shi Xu[1, 2,*], Chuan-Feng Li[1, 2,*] and Guang-Can Guo[1,2]

[1]*CAS Key Laboratory of Quantum Information, University of Science and Technology of China, Hefei, Anhui 230026, People's Republic of China*

[2] *CAS Center for Excellence in Quantum Information and Quantum Physics, University of Science and Technology of China, Hefei, Anhui 230026, People's Republic of China*

[3] *Accelerator Laboratory, School of Physics and Technology, Wuhan University, Wuhan, Hubei 430072, People's Republic of China*

*Corresponding author: jsxu@ustc.edu.cn, cfli@ustc.edu.cn

**Abstract**

**Defects in silicon carbide have been explored as promising spin systems in quantum technologies. However, for practical quantum metrology and quantum communication, it is critical to achieve the on-demand shallow spin-defect generation. In this work, we present the generation and characterization of shallow silicon vacancies in silicon carbide by using different implanted ions and annealing conditions. The conversion efficiency of silicon vacancy of helium ions is shown to be higher than that by carbon and hydrogen ions in a wide implanted fluence range. Furthermore, after optimizing annealing conditions, the conversion efficiency can be increased more than 2 times. Due to the high density of the generated ensemble defects, the sensitivity to sense a static magnetic field can be research as high as $\eta_B \approx 11.9 \mu T/\sqrt{Hz}$, which is about 15 times higher than previous results. By carefully optimizing implanted conditions, we further show that a single silicon vacancy array can be generated with about 80 % conversion**

**efficiency, which reaches the highest conversion yield in solid state systems. The results pave the way for using on-demand generated shallow silicon vacancy for quantum information processing and quantum photonics.**

**Keyword:**

**Silicon carbide, silicon vacancy, implantation, magnetic sensing, single photon sources**

In recent years, color centers in silicon carbide (SiC) have been demonstrated as promising physical platforms for quantum science[1-11]. SiC is a well-known semiconductor material which has wide applications in high-power and high-temperature electronic devices. Moreover, SiC has technological advantages due to the well-developed device fabrication protocols and inch-scale growth. Besides some bright single photon emitters[3-7], SiC also has two types of defect spins, including the silicon vacancy and divacancy defects[1,2,8-11]. Similar with nitrogen-vacancy (NV) centers in diamond[12], these spins can be polarized by optics and manipulated by microwaves at room temperature (RT). Moreover, their photoluminescence (PL) spectrum are in the near infrared, which would have weaker scattering losses at interfaces and signal attenuation in optical fibers[12] than that of NV centers in diamond. Recently, silicon vacancy ($V_{Si}$) defects in 4H-SiC stand out as a favorable system for the quantum technology including quantum information process and quantum sensing, due to their unique properties such as photostability, half-integer $S = 3/2$ spin and long spin coherence time at RT [9-11,13,14,16-21].

For practical quantum sensing and quantum communication, it is critical to generate shallow $V_{Si}$ defects in 4H-SiC with high enough efficiency. Previously, there are two methods to generate shallow $V_{Si}$ defects: carbon ion implantation[20] and focus silicon ions beam[21], etc. However, the conversation efficiencies are less than 20 % and the generation effect of $V_{Si}$ defects using other different implanted ions is still little known[20,21]. Since the ions implantation create residual radiation damage, it would degrade the coherence properties of $V_{Si}$ defects to an extent that is hardly usable as a spin probe[2,22-24]. Moreover, due to the low counts of $V_{Si}$ defects (about 10 kcps), in order to conveniently integrate with photonic devices[25,26], it is necessary to further improve the conversation efficiency to on-demand generate single $V_{Si}$ defects. Furthermore, efficiently generate high density $V_{Si}$ defect ensembles will tremendously increase the sensitivity in quantum sensing applications and be useful in the investigation of many-body dynamics with defect ensemble interactions[16-19, 26,27,28].

In this work, we compare the generation effect of shallow $V_{Si}$ defects in 4H-SiC by using three different implanted ions with a wide implanted fluence. Through measure the counts and optically detected magnetic resonance (ODMR) spectrum, the implanted effect by helium ions is shown to be better than that by carbon and hydrogen ions in a wide fluence. Moreover, we also optimize the annealing conditions for the three different implanted ions. The conversation effect is found to be increased more than 2 times at the optimal annealing conditions. The ODMR signal of the generated high density shallow $V_{Si}$ defect ensembles are measured with different external magnetic field with the magnetic sensitivity detected to be about $11.9\ \mu T/\sqrt{Hz}$, which is about 15 times higher than previous results. Finally, we generate a shallow single $V_{Si}$ defect array in SiC with a high implanted conversation efficiency of about 80 %. Our on-demand generation of single and high density ensemble of shallow $V_{Si}$ defects would be directly used for defect based quantum photonics and quantum sensing with technological materials.

In the experiment, we use a commercially available high-purity 4H-SiC epitaxy layer (thickness is about 7 μm, Power way wafer) sample[20,21]. In order to compare the implanted effect of different ions, we generate the shallow $V_{Si}$ defects in the sample by implanting the hydrogen ($H_2^+$), helium ($He^+$) and carbon ($C^+$) with the same energy (40 keV for $H_2^+$, 20 keV for $He^+$ and $C^+$), and the influence ranging from $1\times 10^{11}$ cm$^{-2}$ to $1\times 10^{14}$ cm$^{-2}$, respectively. For the demonstration of generated single silicon vacancy array, we use the electron-beam lithography (EBL) to make a 50-nm-diameter arrays ($2\times 2$ μm$^2$) on a 200 nm-PMMA layer, which is deposited on the SiC surface[20]. Since the 20 keV helium and hydrogen implanted ions can penetrate through the 200 nm-PMMA layer inferred from the Stopping and Range of Ions in Matter (SRIM) simulation[20,21], we choose to show the case implanted by carbon ions that can be blocked by the PMMA layer. The sample is then implanted by the 20-keV carbon ions with the fluence of $1.3 \times 10^{11}$ cm$^{-2}$ to prepare the single shallow $V_{Si}$ defect array[20]. The emission and spin properties of the defects are characterized by using home build confocal microscopy combined with a microwave system[20,21]. A 730 nm laser is used to excite the $V_{Si}$ defects through an objective. For investigating the defects at RT, we use a 1.3 N. A. oil objective (Nikon) to focus on the sample and collect the fluorescence directly by two avalanche photodiodes (APDs) with a 40-μm-diameter pinhole[13,20,21]. For detecting the low temperature (LT) PL spectrum, we use a Montana Cryostation (4 – 350 K) combined with a confocal system with an infrared 0.65 N. A. objective (Olympus).

With the setup described, we first characterize the PL spectrum and ODMR signal of the generated shallow $V_{Si}$ defects by implanting hydrogen, helium and carbon ions with the dose ranging from $1 \times 10^{11}$ cm$^{-2}$ to $1 \times 10^{14}$ cm$^{-2}$ with the same energy (20 keV). Figure 1a shows the theoretical depth profiles of the generated shallow $V_{Si}$ defects simulated by the SRIM[20,21]. All the depth of the $V_{Si}$ defects are less than 200 nm, and particularly, the depth of defects by the carbon implantation is less than 60 nm, which implies that the generated $V_{Si}$ defects are shallow. The comparison of the RT PL spectrum of the defects generated by different implanted ions with the same dose of $1 \times 10^{13}$ cm$^{-2}$ are shown in Figure 1b. The wavelengths of the PL spectrum are all ranged from 850 to 1050 nm, which are consist with the room-temperature PL spectrum of the $V_{Si}$ defects as measured in previous works[9,13]. Moreover, we can see that, the sample implanted by helium ions has the largest PL intensity.

The LT (5 K) PL spectrum of the generated defects are further measured using the LT confocal system. Two representative LT PL spectrum of the defects implanted by He-$1 \times 10^{13}$ cm$^{-2}$ and He-$1 \times 10^{14}$ cm$^{-2}$ are shown in Figure 1c. Two characteristic peaks at 861.3 nm and 916.0 nm are denoted, which are the zero phonon lines (ZPLs) of the two nonequivalent lattice cites of V1 and V2 centers of $V_{Si}$ defects, respectively[14,20,21]. The inset is the zoom of the ZPLs of V2 defect. The blue lines are the fits to the ZPL data. The full width at half maximum (FWHM) of these two ZPLs are deduced from the fits to be 0.39 ± 0.01 nm and 1.04 ± 0.09 nm for the implanted dose of $1\times10^{13}$ cm$^{-2}$ and $1\times10^{14}$ cm$^{-2}$, respectively. Since the ZPL width is vital to observe the two-photon interference and remote entanglement[30,31], we compare the ZPL FWHM of the generated $V_{Si}$ defects as the implanted fluence. The result is shown in Figure 1d, and the width slightly increases as the fluence increases form $1\times10^{11}$ cm$^{-2}$ (0.18 nm) to $1\times10^{13}$ cm$^{-2}$ (0.39 nm), but then dramatically increases to 1.04 nm for the fluence of $1\times10^{14}$ cm$^{-2}$. The broadening of the width of ZPL may stem from the increasing impurities interactions with the $V_{Si}$ defects[31].

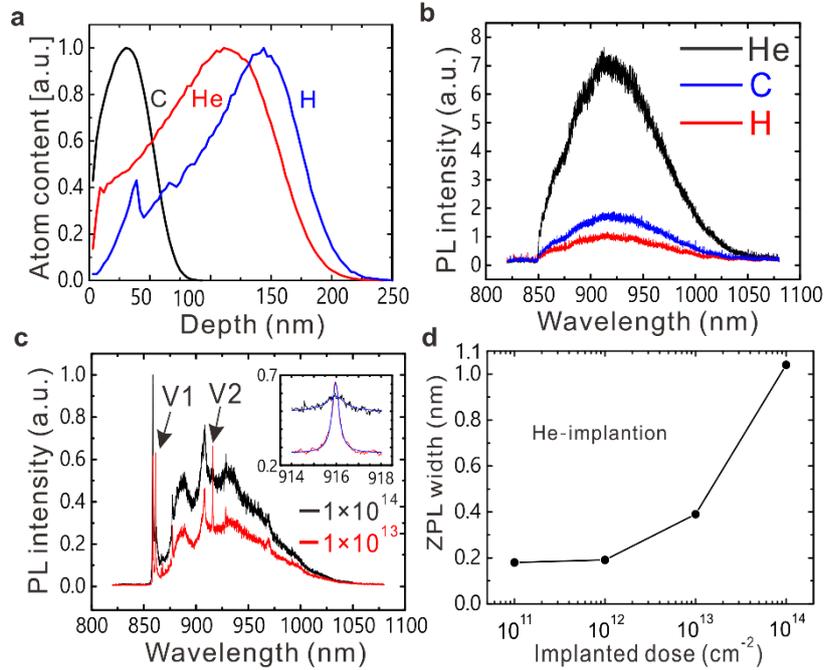

**Figure 1.** Character of the PL spectrum of the implanted shallow $V_{Si}$ defects. (a) The SRIM simulation of depth profiles of the generated shallow $V_{Si}$ defects in 4H-SiC implanted by three types of ions with the same energy (20 keV): hydrogen (H$^+$), helium (He$^+$) and carbon (C$^+$). (b) Comparison of room temperature PL spectrum of implanted $V_{Si}$ defects by the three ions with the same fluence of $1\times10^{13}$/cm$^2$. (c) The low temperature PL spectrum of implanted $V_{Si}$ defects by helium with the fluence of $1\times10^{13}$ cm$^{-2}$ and $1\times10^{14}$ cm$^{-2}$, respectively. The inset is the zoom of the ZPLs of V2 defect. The blue lines are the fits to the ZPLs data. (d) The ZPL width of V2 defects increases as the implanted dose increases.

In order to compare the fluence effect of different implanted ions, we comprehensively investigate the PL intensity of the generated $V_{Si}$ defects with an excitation power of 0.15 mW. Three representative confocal fluorescence images of the defects implanted by C-$1\times10^{13}$ cm$^{-2}$, He-$1\times10^{13}$ cm$^{-2}$ and He-$1\times10^{14}$ cm$^{-2}$ are shown in Figures 2a, 2b and 2c, respectively. The results show that the PL intensity of $V_{Si}$ defects implanted by He is about 4 times larger than that of C with the same dose of $1\times10^{13}$ cm$^{-2}$. The PL intensity of $V_{Si}$ defects implanted by He with a dose of $1\times10^{14}$ cm$^{-2}$ is about 1.7 times larger than that with the dose of $1\times10^{13}$ cm$^{-2}$.

We further compare the mean counts of the shallow $V_{Si}$ defects generated by three different implanted ions with different implanted dose in Figure 2d. The mean counts are calculated as the average counts of the scanned $10\times10$ μm$^2$ areas. It can be seen that, the mean counts of defects for all the three implanted ions increase almost linearly with the fluence ranging from $1\times10^{11}$ cm$^{-2}$ to $1\times10^{13}$ cm$^{-2}$. When the fluence is up to $1\times10^{14}$ cm$^{-2}$, the counts of hydrogen ions implantation still increase linearly, and helium ions implantation increases slowly. However, for the carbon ions implantation, the counts are shown to be a little decrease, which is similar to that observed in the NV centers in diamond implanted by nitrogen ions[29]. The reason for the decrease might due to the ion-induced damage of the crystal lattices which leads to the amorphization of the SiC[29]. The implanted effect of helium is shown to be better than that of hydrogen and carbon for the fluence ranging from $1\times10^{11}$ cm$^{-2}$ to $1\times10^{14}$ cm$^{-2}$, which has the highest conversation efficiency to generated $V_{Si}$ defects. Particularly, the mean counts of the helium implantation are about 2 and 4 times larger than that of the carbon and hydrogen implantation for the fluence of $1\times10^{11}$ cm$^{-2}$, respectively. While, for the higher fluence of $1\times10^{14}$ cm$^{-2}$, the mean counts of the helium implantation are about 8 and 1.5 times larger than that of the carbon and hydrogen implantation, respectively. In our experiment, the PL intensity are almost 10 times higher than previous maximum implanted results[12], which would be useful for high sensitive quantum sensing based on $V_{Si}$ defect ensembles and investigation of many-body dynamics with defect interactions[16-19,26,27,28].

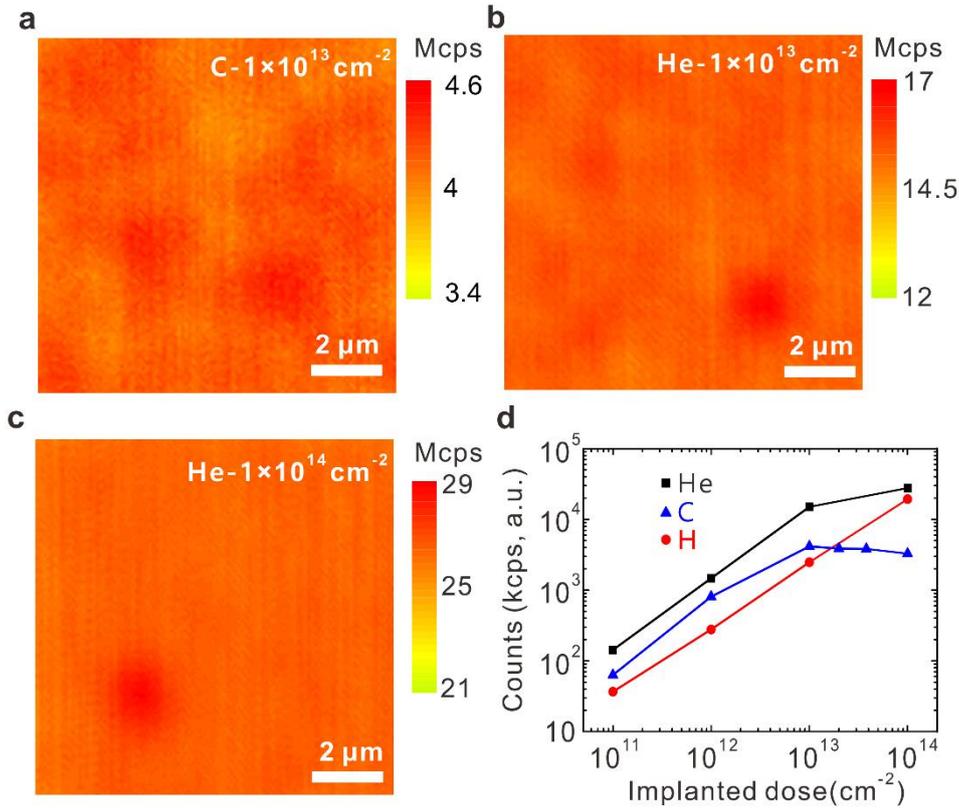

**Figure 2.** PL intensity of defects generated by different implanted ions for different fluence with an excitation power of 0.15 mW. Confocal microscope fluorescence image (10 × 10 μm $^2$) of the implanted shallow $V_{Si}$ defects by different implanted conditions: (a) C-1 × $10^{13}$ cm$^{-2}$, (b) He- 1 × $10^{13}$ cm$^{-2}$ and (c) He-1 × $10^{14}$ cm$^{-2}$, respectively. (d) Comparison of the mean counts of the shallow $V_{Si}$ defects generated by three different implanted ions ($H_2^+$, $He^+$, $C^+$) for different implanted doses.

The annealing has been shown to be an effective method to increase the density of defects[3,13], which is further investigated in our experiment. A glass tube furnace with a high vacuum of 5 × $10^{-4}$ Pa is used to anneal the sample. Figure 3a shows the PL spectrum of $V_{Si}$ defects implanted by He-1 × $10^{13}$ cm$^{-2}$ after different annealing temperature for 0.5 h. It is shown that the PL spectrum profiles keep the same, which implies that the annealing does not change the optical properties of the $V_{Si}$ defects. Moreover, the PL intensity increases with the annealing temperature increasing from 300 °C to 500 °C. However, it decreases after 600 °C annealing, which means that it has an optimal annealing temperature. We then study the effect of annealing on the spin property of the implanted shallow $V_{Si}$ defects. Figure 3b shows the ODMR measurement of the $V_{Si}$ defects (He-$1 \times 10^{13}$ cm$^{-2}$) at zero magnetic field without annealing and after annealing at 500 °C for 1 h, respectively. Red lines are the Lorentzian fits of the ODMR data. We find that the resonant microwave frequencies are almost the same for the samples without annealing (69.5 MHz with a width of 15.6 MHz) and with annealing (70.2 MHz with a width of 15.2 MHz). All the results consist with previous ODMR results of the $V_{Si}$ defects[9,16,20,21], which demonstrate that annealing does not influence the spin property of $V_{Si}$ defects.

Figure 3c shows the normalized intensity of the $V_{Si}$ defects implanted by H-$1 \times 10^{13}$/cm$^2$ as a function of the annealing temperature with annealing time of 0.5 h (black) and 1 h (red), respectively. The mean counts of the scanned $10 \times 10$ μm$^2$ areas before annealing is set as 1. It is shown that the normalized intensity profiles of 0.5 h and 1 h annealing are almost the same, and both of the optimal annealing temperature is 300 °C, and the normalized counts increase about 2 times. As shown in Figures 3d and 3e, both of the normalized intensity profiles of 0.5 h (black) and 1 h (red) annealing for the He-$1 \times 10^{13}$/cm$^2$ and C-$1 \times 10^{13}$/cm$^2$ implantation are also almost the same, and the optimal annealing temperature are 500 °C and 600 °C, respectively. Moreover, the normalized counts increase about 3 and 4 times, respectively. We further investigate the PL enhancement effect of the annealing time at the optimal annealing temperature (C-600 °C, He-500 °C, H-300 °C), which is shown in Figure 3f. The PL enhancement are almost the same for the annealing time from 0.5 h to 4 h, which demonstrates the robustness of the annealing effect.

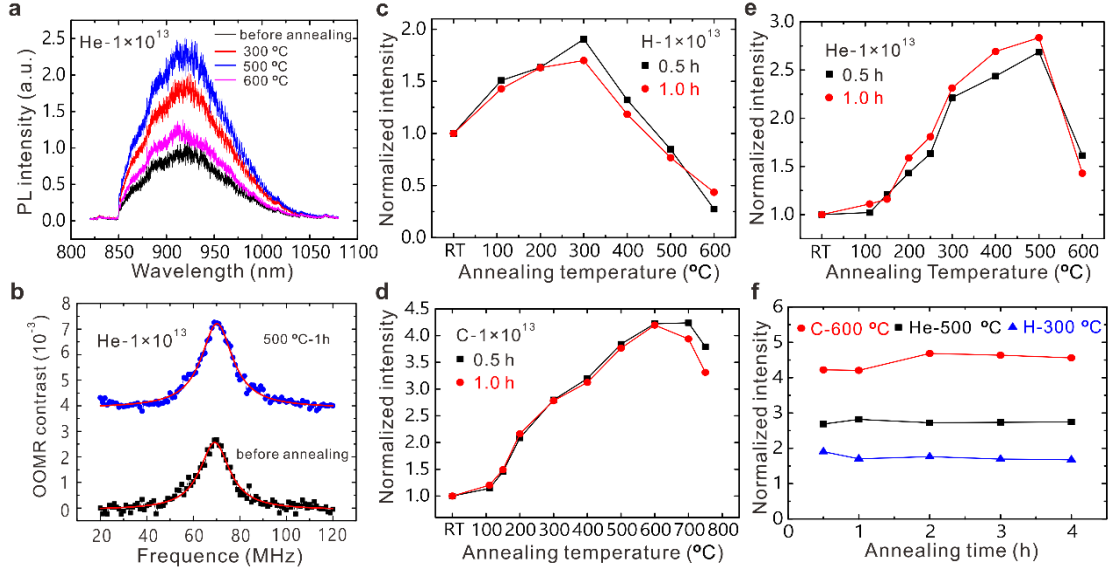

**Figure 3.** PL enhancement after different annealing conditions. (a) The RT PL spectrum of the implanted $V_{Si}$ defects (He-$1\times10^{13}$ cm$^{-2}$) after different annealing temperature with the annealing time of 0.5 h. (b) The ODMR measurement of the implanted shallow $V_{Si}$ defects (He-$1\times10^{13}$ cm$^{-2}$) without annealing and after annealing at 500 °C for 1 h, respectively. The red lines are the Lorentzian fitting of the data. (c)-(e) The normalized intensity of the defects as a function of the annealing temperature with the annealing times of 0.5 h and 1 h for implanted hydrogen, helium and carbon ions ($1\times10^{13}$ cm$^{-2}$), respectively. RT in the X aixs represents the case with no annealing on the sample. (f) Summary of the PL enhancement at corresponding optimized temperature for different annealing time.

Since the $V_{Si}$ defect is a spin qubit and has been used in spin-based quantum information processing and quantum sensing[9,11,13,14,16-20,32,33], we further investigate the spin property of the high concentration shallow $V_{Si}$ defect ensembles (V2 center) (implanted by He-$1\times10^{13}$ cm$^{-2}$, annealing at 500 °C for 1 h) with an external magnetic field. The electronic ground state of the $V_{Si}$ defects is a quartet manifold ($S = 3/2$) and its electronic spin Hamiltonian is:

$$H = D[S_z^2 - S(S+1)/3] + g\,\mu_B B S_z , \qquad (1)$$

where the zero-field-splitting (ZFS) parameters $D$ is 35 MHz, $g = 2$ is the electron g-factor, $\mu_B$ is the Bohr magneton and $B$ is the applied axial static magnetic field. In zero magnetic field, the $|\pm1/2\rangle$ and $|\pm3/2\rangle$ states are both degenerated. Figure 4a shows three ODMR measurement of different magnetic field, with the lowest panel showing the case with zero magnetic field for comparison. By increasing the external axial magnetic field *B*, the spin degeneracy is lifted up, resulting in four distinct energy eigenvalues $E_{\pm 1/2} = -D \pm g\mu_B B/2$, $E_{\pm 3/2} = D \pm 3g\mu_B B/2$ and two dipole-allowed transitions: $|-1/2\rangle \leftrightarrow |-3/2\rangle$ and $|1/2\rangle \leftrightarrow |3/2\rangle$, respectively. The corresponding transition frequencies are $\nu_1 = |2D - g\mu_B B|$ and $\nu_2 = 2D + g\mu_B B$, respectively. When *B* is less than $2D/g\mu_B = 25G$, $\nu_1 = 2D - g\mu_B B$. As shown in Figure 4a (the middle panel), the two resonance frequencies (width) are $\nu_1 =$ *42.2 MHz (10.9 MHz)* and $\nu_2 =$ *98.1 MHz (15.2 MHz)* with the magnetic field at about 10 G. However, when the *B* is larger than 25 G, $\nu_1 = g\mu_B B - 2D$. As shown in Figure 4a (the upper panel), the two transition frequencies (width) are $\nu_1 =$ *41.1* MHz *(5.5 MHz)* and $\nu_2 =$ *179.2 MHz (11.6 MHz)* with the magnetic field at about 39 G.

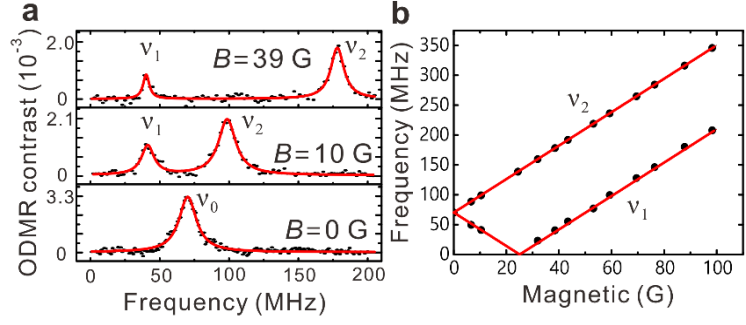

**Figure 4.** ODMR measurement with different external magnetic field at room temperature. (a) The ODMR measurement of the V$_{Si}$ defects (He-1 1013 cm-2, annealing at 500 °C for 1 h) for three different axial magnetic field. The red lines are the Lorentzian fit for the ODMR data. (b) The resonance frequencies of the ODMR signal as a function of the axial magnetic field from 0 G to 100 G. The red lines are the theoretical predictions using the Hamiltonian of the V$_{Si}$ defects.

The transition frequencies of the ODMR signal as a function of the axial magnetic field from 0 G to 100 G is shown in Figure 4b. The red lines are the calculations for the dipole-allowed two transitions frequencies using the Hamiltonian of the V$_{Si}$ defects. The experiment ODMR spectrum is in good agreement with theoretical predictions using equation (1). To evaluate the $V_{Si}$ defect-based magnetic sensing sensitivity, the shot-noise-limited DC magnetic field sensitivity $\eta_B$ is approximated through the following equation $\eta_B \approx \frac{h}{g\mu_B} \frac{\Delta\nu}{C\sqrt{R}}$, where the $h$ is the Planck constant, $R$ is the rate of detected photons, $C$ is the contrast of the ODMR, and $\Delta\nu$ is the ODMR width [34,35]. In the experiment, the mean ODMR width $\Delta\nu = $ *11.7 MHz*, the contrast of the ODMR $C$ is about 0.2 %, and the maximum detected photons (saturated count) $R$ is about *310 Mcps*, we can then deduce the sensitivity $\eta_B \approx 11.9 \mu T/\sqrt{Hz}$, which is about 15 times higher than previous results[17]. It can be expected that the ODMR width can be further reduced to be *0.5 MHz* using the pulse ODMR, and the corresponding magnetic sensitivity can be further improved to about *500* $nT/\sqrt{Hz}$ [17]. Moreover, the magnetic sensitivity can be further improved by using the isotopically purified SiC substrate[36].

Finally, we character the generate efficiency of the carbon implanted shallow single $V_{Si}$ defect arrays. As a contrast, we investigate two samples, one is as implanted, another is optimal annealing (600 °C for 1 h). Figures 5a and 5d are two representative confocal fluorescence images (20×20 μm$^2$) of the implanted shallow single $V_{Si}$ defect arrays before and after annealing with an excitation power of 0.5 mW, respectively. It was shown that in both conditions, they display clear arrays with the prominent increase of the concentration in the annealing sample. In order to identify a single $V_{Si}$ defect, we measure the corresponding second-order correlation function of the defect in the implanted aperture denoted by the white circle in Figures 5a and 5b, respectively. Since the fluorescent count rate of the single $V_{Si}$ defect is low (about 12 *kcps*, data not shown), the background fluorescence has been included in the evaluation of $g^2(\tau)$. The auto-correlation function $g^2(t)$ is corrected from normalized raw data $C_N(t)$ using the function $g^2(\tau) = (C_N(\tau)-(1-\rho^2))/\rho^2$, where $\rho = s/(s+b)$, and $s$ and $b$ are the signal and background counts, respectively[8,13,20,21]. The background-corrected $g^2(\tau)$ are shown in Figures 5b and 5e, respectively. The red lines are the fits according to the function $g^2(\tau)$ = *1-(1+a)\*exp(-|τ|/τ$_1$)+a\*exp(-|τ|/τ$_2$)*, where *a*, *τ$_1$*, *τ$_2$* are fitting parameters. It can be seen that, in both cases, g$^2$(*0*) is close to 0, which demonstrate they are single $V_{Si}$ defects.

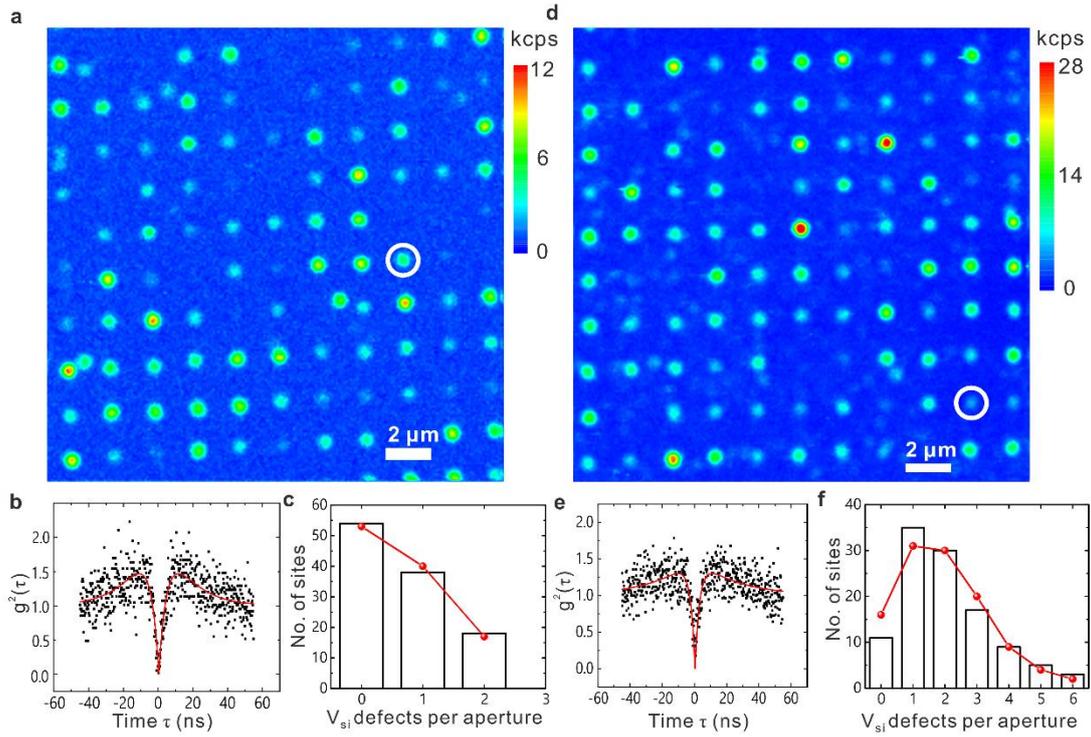

**Figure 5.** Character of the carbon implanted shallow single $V_{Si}$ defect arrays. (a) and (d) are two representative confocal fluorescence images (20 × 20 μm$^2$) of the implanted shallow single $V_{Si}$ defect arrays before and after annealing (600 °C for 1 h) with an excitation power of 0.5 mW, respectively. The scale bars are 2 μm. (b) and (e) are the second-order correlation function measurement of the single $V_{Si}$ defects (correspond to the circled single $V_{Si}$ defects in (a) and (b), respectively), respectively. The red lines are the fits of the data. (c) and (f) are the statistics of the number of $V_{Si}$ defects per implanted aperture. The data are fitted with the Poisson distribution (red curves).

Then we count the number of $V_{Si}$ defects per implanted aperture for 110 apertures in both samples and the corresponding results are shown in Figures 5c and 5f, respectively. The data are fitted using Poisson distribution (red curves). Inferred from the fits, the average number of $V_{Si}$ defects formed per aperture are 0.75 ± 0.05 and 1.94 ± 0.12, respectively. Since the implanted influence corresponds to about 2.5 carbon atoms per aperture, the conversion yield of the implanted carbon ions into the $V_{Si}$ defects are about 30% ± 2% and 78% ± 5%, respectively. The high conversion yield (about 80%) is about 4 times higher than pervious results[20], which reaches the highest conversion yield in solid state systems[37].

In summary, we provide a method to on-demand generate shallow single $V_{Si}$ defect arrays with a high conversion efficiency of about 80% and high concentration defect ensembles in 4H-SiC. The comparison of implanted results demonstrates that the effect of helium implantation are better than that by hydrogen and carbon implantation from a low fluence ($1 \times 10^{11}$ cm$^{-2}$) to a high fluence ($1 \times 10^{14}$ cm$^{-2}$). Through optimize annealing temperature and annealing time, the PL intensity increases about more than 2 times. Moreover, using the implanted high concentration ensemble $V_{Si}$ defects, the magnetic sensing sensitivity can be increased to be $\eta_B \approx 11.9 \mu T / \sqrt{Hz}$, which is about 15 times higher than previous results[17]. The magnetic sensitivity can be further improved by using the pulse ODMR[17] and the isotopically purified SiC substrate[36].

Our experimental results open up several practical applications. Firstly, the high conversation efficiency of single $V_{Si}$ defects would lead to little residual radiation damage[22], which can prolong the coherence time of defect spins and is useful in quantum information processing[14] and quantum sensing[16,17]. Moreover, it will also be good for coupling $V_{Si}$ defects to photonics devices[9,25,26]. Secondly, the high concentration $V_{Si}$ defect ensembles can be used to achieve highly sensitive magnetic and temperature sensing[16-19,38-40], and study many-body dynamics with defect interactions[26,27,28]. Using the laser ablated[41] or reactive ion etching[42], we can make high quality nano-SiC particles with the implanted sample, which would lead to high sensitivity, nanoscale spatial resolution sensor with photostability and chemical inertness. Finally, the methods may be used to generate other types of defects in SiC with a high quality, such as dicacancy[1,2].

**Acknowledge**

J. F. Wang and Q. Li contributed equally to this work. We thank Prof. Liping Guo for his help with the implantation. This work was supported by the National Key Research and Development Program of China (Grant No. 2016YFA0302700), the National Natural Science Foundation of China (Grants No. 61725504, 61327901, 61490711, 11821404 and 11774335), the Key Research Program of Frontier Sciences, Chinese Academy of Sciences (CAS) (Grant No. QYZDY-SSW-SLH003), Anhui Initiative in Quantum Information Technologies (AHY060300 and AHY020100), the Fundamental Research Funds for the Central Universities (Grant No. WK2470000020 and WK2470000026). The ion implantations were carried out on the implanter (JZM 5900) at the acelerator laboratory of Wuhan University. This work was partially carried out at the USTC Center for Micro and Nanoscale Research and Fabrication.